\documentclass[preprint,review,11pt]{elsarticle}

\usepackage{amssymb}
\usepackage{multirow}
\usepackage{amsmath}

\usepackage{algcompatible}
\usepackage{algorithm}

\usepackage{xcolor}

\journal{Journal of Parallel and Distributed Computing}

\def\etal{\emph{et al.}}
\def\eg{\emph{e.g.}}

\begin{document}

\begin{frontmatter}



%\title{Cloud-FedLS: a Framework for Managing Federated Learning Applications on Cloud Environments}
\title{Multi-FedLS: a Framework for %Managing 
Cross-Silo Federated Learning Applications on Multi-Cloud Environments}

\author[uff,su]{Rafaela C. Brum}
\ead{rafaelabrum@id.uff.br}
\author[uerj]{Maria Clicia Stelling de Castro}
\ead{clicia@ime.uerj.br}
\author[su]{Luciana Arantes}
\ead{luciana.arantes@lip6.fr}
\author[uff]{L\'{u}cia Maria de A. Drummond}
\ead{lucia@ic.uff.br}
\author[su]{Pierre Sens}
\ead{pierre.sens@lip6.fr}

\affiliation[uff]{organization={Fluminense Federal University},
            addressline={Av. Gal. Milton Tavares de Souza, s/nº, São Domingos},
            city={Niterói},
            state={Rio de Janeiro},
            country={Brazil}}
\affiliation[su]{organization={Sorbonne Université, LIP6, CNRS},
            addressline={4 Place Jussieu},
            postcode={75005},
            city={Paris},
            country={France}}
\affiliation[uerj]{organization={State University of Rio de Janeiro},
            addressline={Rua São Francisco Xavier, 524, Maracanã},
            city={Rio de Janeiro},
            state={Rio de Janeiro},
            country={Brazil}}

\begin{abstract}

% Rafaela: Revisar por último
Federated Learning (FL) is a distributed Machine Learning (ML) technique that can benefit from cloud environments while preserving data privacy. We propose Multi-FedLS, a framework that manages multi-cloud resources, reducing execution time and financial costs of Cross-Silo Federated Learning applications by using preemptible VMs, cheaper than on demand ones but that can be revoked at any time. Our framework encloses four modules: Pre-Scheduling, Initial Mapping, Fault Tolerance, and Dynamic Scheduler.   
This paper extends our previous work \cite{brum2022sbac} by formally describing the Multi-FedLS resource manager framework and its modules.
% , focusing on the last two. 
%We execute
Experiments were conducted with three Cross-Silo FL applications on CloudLab and %we present 
a proof-of-concept confirms that Multi-FedLS %manages the execution in
can be executed on a multi-cloud composed by AWS and GCP, two commercial cloud providers. Results show that the problem of executing Cross-Silo FL applications in multi-cloud environments with preemptible VMs can be efficiently resolved using a mathematical formulation, fault tolerance techniques, and a simple heuristic to choose a new VM in case of revocation.
% XXXXX SO esse resultado??? Nao pode ser!!! Isso é obvio! \textcolor{blue} {Our results show that fault-tolerant strategies are crucial to efficiently executing an FL application in a multi-cloud environment ISSO NAO È RESULTADDO DE UM TRABALHO DESSE TAMANHO!!! DE JOURNAL! }. tem que dizer que os testes mostram q o framework se mostra uma boa alternativa para aplicacoes tal e tal...  nos cenarios assim e assim permitindo que problemas importantes como ... possam ser resolvidos de forma eficiente... etc...
\end{abstract}


% \begin{highlights}
% \item We formally describe Multi-FedLS, a multi-cloud resource manager for Cross-Silo FL applications. It takes advantage of preemptible VMs, which are cheaper resources, but can be revoked at any time. 
% \item We execute three Cross-Silo FL applications in Multi-FedLS, being two FL benchmarks and one real-world application.
% \item We also present a proof of concept that our resource manager works in a real multi-cloud scenario.
% \end{highlights}


\end{frontmatter}

\section{Introduction}

Proposed by McMahan~\etal~\cite{mcmahan2017communication}, Federated Learning (FL) is a distributed machine learning (ML) technique where the participants (clients) collaboratively train a ML model without sharing their respective private data~\cite{shen2021difference_distributed_federated_learning}. 
Since there is an increasing concern with data privacy, particularly with current data protection laws (\eg~GPDR\footnote{https://gdpr-info.eu/} in Europe), FL is an extremely interesting and promising approach. 

Denoted Model-Centric Federated Learning~\cite{yang2019federated}, FL has a client-server architecture where each client trains the model locally and communicates only the model weights to the central server which keeps the current global model. The server is responsible for receiving all client weights, updating the global model, and communicating the new model weights to the clients.

The model-centric FL can be classified into two types: {\it Cross-Device} FL, composed of a large number of clients which are low-powered devices, like mobile phones~\cite{mcmahan2017communication} or IoT devices~\cite{nguyen2021iot}, and {\it Cross-Silo} FL, composed of a few but powerful clients, usually companies or institutions, with similar datasets willing to create a central model with private large dataset repositories, denoted silos.  
Note that in Cross-Device FL, the set of clients can strongly vary over the execution, while in Cross-Silo FL, all clients can be considered to be available during the whole execution since they are machines with high-performance capabilities or even clusters.

In Cross-Silo FL applications, the scheduling and resource provisioning problems can be related to the scheduling of distributed Machine Learning (ML) applications~\cite{mohamadi2019distributed_ml_stragglers, liu2020distributed_ml_networks, yu2021distributed_ml_locality} as there is the assumption that the clients' jobs are always available.   
On the other hand,  
it is worth pointing out that there are two main differences between scheduling distributed ML and FL: (i) in distributed ML, the scheduler assumes that all clients have the same amount of data, and (ii) clients can share data among themselves.

Training ML algorithms require an enormous amount of data, often stored using cloud storage services~\cite{li2010cloudcmp, huan2013big_data_storage, leavitt2013storage}. These services are part of the Infrastructure-as-a-Service (IaaS) model offered by  a cloud provider and they can vary in performance and costs. Usually, the IaaS services include storage, Virtual Machines (VMs) with different accelerators, and different network bandwidths depending on the cloud provider.

This work focuses on Cross-Silo Federated Learning for multi-cloud platforms where clients' silos are kept in different storage repositories, which provide data privacy guarantees and availability.

A multi-cloud platform is composed of one or more cloud providers that offer multiple choices of virtual machine types in independent and isolated geographic regions\footnote{https://aws.amazon.com/about-aws/global-infrastructure/}$^{,}$\footnote{https://cloud.google.com/about/locations}, with different execution and communication times and costs. %Therefore, multi-cloud platforms require an adequate choice of cloud resource allocation and are critical to obtaining an efficient execution of FL applications. 
Hence, in order to have  efficient executions of FL applications on a  multi-cloud platform, the good choice of cloud resource allocation is imperative.
Wrong ones may increase execution cost and time~\cite{ostermann2010performance, kotas2018comparison}. 
%So to avoid losses, a multi-cloud environment should treat the configuring problems in clouds, FL execution monitoring, and a new VM selection whenever one of the currently allocated ones becomes unavailable.

Therefore, assuming datasets previously allocated in one or more storage systems of different cloud providers, 
the question is how to select the most suitable virtual machines to execute the clients and the server in a multi-cloud platform in order to minimize execution time and financial costs.

%This work focuses on Cross-Silo Federated Learning, where clients keep their silos in diverse cloud providers with different storage repositories, which offer data privacy guarantees and availability. The question is how to select the best virtual machines to execute the clients and the server in a multi-cloud platform, assuming datasets previously allocated in one or more storage systems belonging to different cloud providers, minimizing execution time and financial costs.

%Aiming at solving this resource management problem, we propose in this paper the {\it Multi-FedLS} framework. It uses the open source tool Flower \cite{beutel2021flower} and encloses four modules: Pre-Scheduling, Initial Mapping, Fault Tolerance, and Dynamic Scheduler, depicted in Section~\ref{sec:FLarch}. 

%Two of our previous works~\cite{brum2022wcc, brum2022sbac} have described  and evaluated the Pre-Scheduling and Initial Mapping modules. 
%In these works, 
%The obtained results
Aiming at solving this resource management problem, two of our previous works~\cite{brum2022wcc, brum2022sbac} proposed the multi-cloud framework {\it Multi-FedLS} which is composed by the Pre-Scheduling and Initial Mapping modules. The Pre-Scheduling module executes several tests with a dummy application to obtain slowdown metrics, which are then used by the Initial Mapping module to define a first static scheduling of the tasks. 
%
%Experiments with {\it Multi-FedLS} were conducted on a Google and AWS multi-cloud platform by varying the number of clients and each client's dataset location  %Furthermore, %these works as well as the type of VMs and bounds of the maximum number of VMs per region. 
%One constraint was a limit of 4 VMs with GPU gave us by the real cloud providers, Amazon Web Services (AWS) and Google Cloud Provider (GCP). 

The above works considered that {\it Multi-FedLS} consists of one or more cloud providers whose VMs are never revoked neither fail. However, in order to reduce monetary costs, the allocation of spot (preemptible) VMs should be considered since they are offered with a high discount. On the other hand, such VMs can be revoked at any time by the respective cloud provider. Since an FL application is a long-running one, {\it Multi-FedLS} should %offer a mechanism to 
(1) periodically save the computed weights of both the server and clients in a stable storage and (2) be able to restart the weight computing task
%as well as to restart them 
in a new VM, if the one in which the task executes revokes or fails.
%previous VM executing them are revoked or fail. 
Considering these two features, the current work extends the {\it Multi-FedLS} framework presented in~\cite{brum2022wcc, brum2022sbac}  by adding to it two new modules: the Fault Tolerance and the Dynamic Scheduler. The former implements a checkpoint mechanism while the latter reconfigures the Initial Mapping, also choosing a new VM to replace the revoked or failed one and restarting the corresponding task. %faulty task.

For the sake of scalability, performance experiments were carried out on CloudLab~\cite{cloudlab_paper}, a platform that provides different types of machines in various U.S. states, emulating a multi-cloud environment and then on a real two-cloud platform composed of Amazon Web Services (AWS) and Google Cloud Provider (GCP). 
We have evaluated {\it Multi-FedLS} with  
a use-case application, which classifies tomography images into with or without tumor lymphocytes, and two FL benchmark datasets.

The paper is organized as follows. Section \ref{sec:relworks} introduces the related work. Section~\ref{sec:app_models} describes the application and environment models used in {\it Multi-FedLS}. Section~\ref{sec:FLarch} shows the {\it Multi-FedLS} architecture using a Flower-based framework.
%The experiment setup and discussions are the focus of Section~\ref{sec:experiments}. %Finally,
Results from experiments on both CloudLab and the AWS-GCP environment are presented and discussed in Section~\ref{sec:experiments}.
Section \ref{sec:conclusion} concludes the work and proposes some future research directions.

\section{Related Work}
\label{sec:relworks}

Since the first FL algorithm~\cite{mcmahan2017communication}, many resource management optimizations, such as the selection of just a fraction of client tasks in each scheduling communication round~\cite{yang2020scheduling, buyukates2021timely_communication, pilla2021optimal}, or new aggregation strategies~\cite{li2018federated, abdelmoniem2023comprehensive} have been proposed. However, they are effective for Cross-Device FL which has a huge number of clients. 

To the best of our knowledge, there are few papers on Cross-Silo FL and none of them focus on resource management of tasks but just on FL solutions specific to given applications~\cite{rajendran2021cloud, huang2021personalized_cross_silo, li2021federated}.

% NOVO
There exist several works about resource management on clouds (using single or multi-clouds)~\cite{shastri2018cloud, varshney2019autobot, javier2019reducing, Chhabra2022, yin2022stochastic, zhou2022farspot, teylo2023scheduling, Karaja2023, Brotcorne2023resource_provisioning}.
Some propose using spot/preemptible VMs to reduce costs and handle possible revocations of these VMs~\cite{shastri2018cloud, varshney2019autobot, javier2019reducing, zhou2022farspot, teylo2023scheduling}. However, most of them focus on Bag-of-Tasks applications, and others consider communication between tasks in their solution~\cite{shastri2018cloud, zhou2022farspot} using only AWS as the cloud provider and do not assume any synchronization barrier.

Our scheduling problem consists in finding the optimal placement for each client and the server of an FL application, which has synchronization barriers along the execution, aiming at minimizing the execution and financial costs. 
Since our proposal is not tailored for any particular Cross-Silo FL application, it does not explore server aggregation approaches for a specific ML model.

Based on the above points, works on scheduling distributed ML applications, especially those that take into account stragglers~\cite{mohamadi2019distributed_ml_stragglers}, network performance~\cite{liu2020distributed_ml_networks}, or locality awareness~\cite{yu2021distributed_ml_locality} are related to ours.

During the distributed ML training, Amiri~\etal~consider in~\cite{mohamadi2019distributed_ml_stragglers} that, among the workers,  there exist transient stragglers. 
To handle them, the authors divide the central dataset into $n$ small chunks, assuming that each worker will compute over a number of them sequentially. Then, the 
server sends each piece to multiple workers and waits for $k$ results from different chunks, $k \leq n$. 
We should point out that in FL scenarios, the data exchange between tasks is prohibitive to preserve privacy.

Liu~\etal~consider switches that need to reconfigure themselves to send data in both directions saving energy. 
% Thus, t
They proposed two scheduling policies for distributed ML training jobs execution. Besides handling possible network contention, the authors assume that all workers have the same dataset size and computational power, which are not always the case in FL scenarios.

Respecting some resource and locality-aware constraints, Yu~\etal~\cite{yu2021distributed_ml_locality} formulate an optimization model which allocates different distributed ML jobs into a set of physical machines. 
Despite the locality-aware scheduling problem, the work has assumptions that cannot be applied to FL scenario: a central server dataset, frequency of communication between tasks much higher than the Cross-Silo FL one, and the co-location of tasks which might help malicious FL clients discover private information of the others~\cite{truong2021privacy_preservation_fl}.

% NOVO
Regarding the execution of FL applications in clouds, there are few works tackling this problem in the literature~\cite{liu2020client, FANG2020101889, rajendran2021cloud}. However, most of them use the cloud as part of their proposed architecture~\cite{liu2020client, FANG2020101889} not as the environment to execute the FL application. The work of Rajendran~\etal~\cite{rajendran2021cloud} is the only one that runs an FL approach in a cloud environment. Their results were implemented models in a scenario with two clients in a simulated environment and on the Microsoft Azure Cloud Databricks~\cite{azure_databricks}, an open-source tool for data engineering and collaborative data science, to exchange the ML model between two institutions. In our work, we not only use clouds as the execution environment but we select the VMs to reduce costs and execution time, handling eventual failures raised by spot ones.

Papers handling distributed Machine Learning on Clouds usually use cloud services as a tool to evaluate their scheduling proposal~\cite{mohamadi2019distributed_ml_stragglers, yu2021distributed_ml_locality} or to verify the central dataset integrity~\cite{zhao2020distributed_ml_integrity_verification}. To the best of our knowledge, there are a few research papers that consider the features of Cloud Computing to propose their solution~\cite{zhang2020machine_learning_volatile, duong_quang2018distributed_ml_iaas, duong2019fc2, wagenlander_2020_spotnik}. 

Zhang~\etal~\cite{zhang2020machine_learning_volatile} consider spot (or preemptible) cloud instances to train distributed ML tasks and study the convergence of classical algorithms to distributed ML training when the number of workers varies. 
Duong and Quang~\cite{duong_quang2018distributed_ml_iaas, duong2019fc2} present FC$^2$, a system to execute distributed ML jobs into on-demand instances of a cloud provider. 
In~\cite{wagenlander_2020_spotnik}, Wagenländer~\etal~propose Spotnik to deal with preemptive VM instance revocations. The authors assume a decentralized training approach in which the workers communicate with each other in a predefined ring pattern to aggregate their results ignoring faults and privacy.  

However, the above papers assume an equal division of the central dataset among workers, using a single region to allocate all cloud resources. %. These facts% 
Such an approach reinforces the point that data privacy and heterogeneity, which are essential in Federated Learning, are not a concern in distributed ML.

\section{Application and Environment models}
\label{sec:app_models}

A Federated Learning application is composed of one server $s$ and a set of clients $C$. It is also divided into well-defined steps called communication rounds (or rounds) with communication barriers between them. Each communication round has two phases: the {\it training} and the {\it evaluation} one. 

The training phase starts by the server $s$ sending a message $s\_msg_{train}$ which has the initial model weights to all clients. After receiving the weights, each client $c_i \in C$ updates its respective local model and starts training it, using its local dataset. When the training finishes, the client sends back a message $c\_msg_{train}$ with the updated model weights. By the time the server receives $c\_msg_{train}$  messages from all clients, it aggregates the weights and then starts the evaluation phase by sending a message $s\_msg_{aggreg}$ with the aggregated weights to all clients. Similarly to the training phase, upon reception of the latter, a client updates its local model. Finally, the client evaluates its local model with its dataset and sends back the ML metrics in a $c\_msg_{test}$ message to the server that, after receiving and aggregating the metrics from all clients, starts a new round.

Regarding the environment, we consider that $P$ is the set of available cloud providers, each 
provider $p_{j} \in P$ 
has a set of regions $R_{j}$, and the  cost $cost\_t_{j}$ (in \$ per GB) to send any message from one VM in $p_{j}$ to any other VM is constant. Typically, a provider $p_j$ offers a limited number of GPUs ($N\_GPU_{j}$) and vCPUs ($N\_CPU_{j}$). Additionally, in each region $r_{jk} \in R_{j}$, the number of available GPUs and vCPUs is restricted ($N\_L\_GPU_{jk}$ and $N\_L\_CPU_{jk}$, respectively). Moreover, each region $r_{jk} \in R_{j}$ contains a set of available VM instance types, $V_{jk}$, that can be executed in it, and each $vm_{jkl} \in V_{jk}$ contains a number of vCPUs, $cpu_{jkl}$, and a number of GPUs, $gpu_{jkl}$, with a fixed cost (in \$ per second) $cost_{jkl}$. 

Note that some variables of the application and environment models were defined in our previous work~\cite{brum2022sbac}.

\section{Multi-FedLS architecture}
\label{sec:FLarch}
This section describes the architecture of Multi-FedLS. It is composed of four modules, described in the following, and uses the FL tool Flower \cite{beutel2021flower}. 
The latter provides a unified API to create Federated Learning applications whose execution can take place in both real-world and simulated scenarios with any ML framework underneath it.
Flower can execute in several environments with different operating systems and hardware settings. Due to its modular structure, the user only needs to implement a few functions for transforming a regular ML application into a federated one.

Our framework takes advantage of the communication rounds of an FL application. In our previous works~\cite{brum2022wcc, brum2022sbac, brum2021wcc, brum2021eradrj, brum2022compas}, we evaluated a use-case application running on two commercial clouds, Amazon Web Services (AWS) and Google Cloud Provider (GCP). We observed that every round, except the first one, has similar execution times.
Thus, we assume that all rounds have similar execution times reducing our problem of scheduling and managing the resources for the whole FL application as a single FL round.

\begin{figure}[h!tpp]
 \centering
 \includegraphics[width=0.85\textwidth]{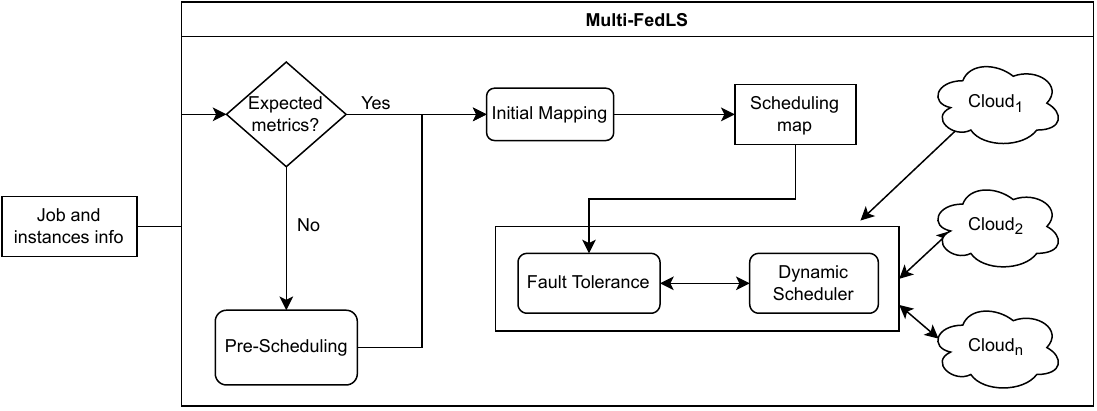}
 \caption{Architecture of Multi-FedLS. Afapted from~\cite{brum2022wcc}}
 \label{fig:framework_design}
\end{figure}

Figure~\ref{fig:framework_design} shows the proposed architecture for the {\it Multi-FedLS} framework. It consists of four main modules: Pre-Scheduling, Initial Mapping, Fault Tolerance with monitoring, and Dynamic Scheduler.
\subsection{Pre-Scheduling module} 

The Pre-Scheduling module executes several tests with a dummy application to obtain two slowdown metrics, %for the Initial Mapping module
as described in previous papers~\cite{brum2022wcc, brum2022sbac}. 
Used as an input for the Initial Mapping, the metrics are: (1) the communication slowdown $sl\_comm_{jklm}$ between every pair of regions $r_{jk}$ and $r_{lm}$ and (2) the execution slowdown $sl\_inst_{jkl}$ in each VM $vm_{jkl}$ of the environment.
%\textcolor{blue}{
% We define slowdown as the ratio between the current pair of regions (or VM) and a chosen baseline pair (or VM).
%}
%
We define communication (resp., execution) slowdown as the ratio between the communication time between a pair of regions (resp., execution time in a VM) and the time in a chosen baseline pair of regions (resp., baseline VM).  

Once the two metrics have been computed, it is not necessary to re-execute the dummy application in every framework execution, but only when there is a change in the regions or the VMs in the environment. 

The baseline values for the current FL job are also computed by the Pre-Scheduling module. These values are (1) the execution time of each client in the baseline VM and (2) the message exchange time of the job model in the baseline pair of regions. The execution time of a client $c_i$ is divided into the training execution time $train\_bl_{i}$ and test execution time $test\_bl_{i}$. The message exchange time is also divided into training and test, where $train\_comm\_bl$ and $test\_comm\_bl$ are the time spent sending messages during the training and the test respectively.

\subsection{Initial Mapping module} \label{sec:initial-mapping}

The Initial Mapping module starts by calculating the expected communication and computation times of the current FL job of all pair of regions and VMs. The communication time $t\_comm_{jklm}$ between regions $r_{jk}$ of provider $p_{j}$ and $r_{lm}$ of provider $p_{l}$ and the computation time $t\_exec_{ijkl}$ of client $c_i$ in VM $vm_{jkl}$ are given by Equation \ref{eq:t_comm} and Equation \ref{eq:t_exec} respectively.
\begin{equation}
\resizebox{0.89\hsize}{!}{$t\_comm_{jklm} = (train\_comm\_bl + test\_comm\_bl) \times sl\_comm_{jklm}$} \label{eq:t_comm}
\end{equation}
\begin{equation}
\resizebox{0.65\hsize}{!}{$t\_exec_{ijkl} = (train\_bl_{i} + test\_bl_{i}) \times sl\_inst_{jkl}$} \label{eq:t_exec}
\end{equation}

The complexity of scheduling tasks in distributed computing resources is proven to be NP-complete~\cite{ullman1975np}, which makes it challenging even in simple scenarios. Furthermore, the features of multi-clouds increase the difficulty to solve the problem. As a result, we have modeled our scheduling problem as a Mixed-Integer Linear Programming problem with two objectives: to minimize the total execution time (makespan) and the monetary cost. Additionally, the scheduling solution should comply with two constraints: the deadline (T) and the budget (B), as specified by the user. 
As we assume that the FL application executes through $n_{rounds}$ with similar execution times, we obtain the
%Given that the FL application runs through $n_rounds$, each having an equivalent duration, it becomes possible to derive the 
maximum budget ($B_{round}$) and deadline ($T_{round}$) of a round %for an individual round 
by dividing $B$ and $T$ by $n_{rounds}$. Consequently, the scheduling problem can be formulated for a single round.

The mathematical formulation, which minimizes both the execution time and the costs of the FL jobs running in the clouds, solved by the Initial Mapping, is detailed in our previous work~\cite{brum2022sbac}. Usually, the most expensive VMs of a cloud take less time to execute. Thus, having conflicting objectives, we apply a user-defined weight value to create a single objective problem. 
% Table \ref{tab:var} summarizes all variables of the framework.
Table \ref{tab:var} summarizes the notation and variables used in \textit{Multi-FedLS}.

\begin{table}[htpb]
\caption{Notation and variables used in our framework.} \label{tab:var}
\begin{center}
\scriptsize
\begin{tabular}{|r|l|}
\hline
\textbf{Name} & \textbf{Description} \\
\hline
$C$ & Set of clients in the FL application \\
\hline
$P$ & Set of available cloud providers \\
\hline
$p_{j}$ & A cloud provider \\
\hline
$R_{j}$ & Set of regions available in provider $p_j$ \\
\hline
$r_{jk}$ & A region of provider $p_{j}$ \\ 
\hline
$V_{jk}$ & Set of instance types available in region $r_{jk}$ \\
\hline
$vm_{jkl}$ & A instance type of region $r_{jk}$ \\
\hline
$B_{round}$ & Budget to a single FL round \\
\hline
$T_{round}$ & Deadline to a single FL round \\
\hline
$T_{max}$ & Time of a single FL round \\
\hline
$N\_GPU_{j}$ & Number of available GPUs in the provider $p_{j}$ \\
\hline
$N\_CPU_{j}$ & Number of available vCPUs in the provider $p_{j}$ \\
\hline
$cost\_t_{j}$ & Cost (in \$ per GB) of sending a message from provider $p_{j}$ \\
\hline
$N\_L\_GPU_{jk}$ & Number of available GPUs in region $r_{jk}$ \\
\hline
$N\_L\_CPU_{jk}$ & Number of available vCPUs in region $r_{jk}$ \\
\hline
$cpu_{jkl}$ & Number of vCPUs in the instance type $vm_{jkl}$ \\
\hline
$gpu_{jkl}$ & Number of GPUs in the instance type $vm_{jkl}$ \\
\hline
$cost_{jkl}$ & Cost (in \$ per second) of instance type $vm_{jkl}$ \\
\hline
$c_{i}$ & A client of the FL application \\
\hline
$size(s\_msg_{train})$ & Size of training message sent by the server to a client \\
\hline
$size(s\_msg_{aggreg})$ & Size of test message sent by the server to a client \\
\hline
$size(c\_msg_{train})$ & Size of training message sent by a client to the server \\
\hline
$size(c\_msg_{test})$ & Size of test message sent by a client to the server \\
\hline
$total\_costs$ & Total financial costs of a single FL round \\
\hline
$t_{m}$ & Total execution time (makespan) of an FL round \\
\hline
$t\_exec_{ijkl}$ & Computational time (training and test) of client $c$ in $vm_{jkl}$ \\
\hline
$t\_comm_{jklm}$ & Communication time (training and test) between regions $r_{jk}$ and $r_{lm}$ \\
\hline
$t\_aggreg_{jkl}$ & Aggregation time of server in $vm_{jkl}$ \\
\hline
$vm\_costs$ & Total financial cost of all VMs in a single FL round \\
\hline
$comm\_costs$ & Total financial cost of message exchange within an FL round \\
\hline
$x_{ijkl}$ & Binary variable which indicates if client $c_{i}$ executes on $vm_{jkl}$ or not \\
\hline
$y_{jkl}$ & Binary variable which indicates if the server executes on $vm_{jkl}$ or not \\
\hline
$\alpha$ & Weight given by the user for the objectives (ranging from 0 to 1) \\
\hline
$T_{max}$ & Maximum possible makespan regarding all clients and possible VMs \\
\hline
$cost_{max}$ & Maximum total cost considering $T_{max}$, all possible locations and VMs \\
\hline
\end{tabular}
\end{center}
\end{table}

The objective function (Equation \ref{eq:obj}) uses the weight $\alpha$, defined by the user, to express how much each objective should be taken into account. The objectives are the makespan of the application, $t_m$, and the execution cost, $total\_costs$. The latter comprises the VMs cost, $vm\_costs$, plus the communication cost, $comm\_costs$.
\begin{equation}
 \min \alpha \times total\_costs + ( 1 - \alpha) \times t_m \label{eq:obj}
\end{equation}

%The VMs execution cost is calculated using Equation \ref{eq:vm_costs}, being $x_{ijkl}$ the binary variable representing whether or not a client $c_i$ executes on VM $vm_{jkl}$ of the region $r_{jk}$ of provider $p_j$, and $y_{jkl}$ the binary variable for the same information for the server $s$. Besides, the communication cost is computed using Equation \ref{eq:transfer_costs}, where $x_{ijkl}$ and $y_{mno}$ has the same meaning as before, and $comm_{jm}$ is the cost to send and receive messages between provider $p_{j}$ and $p_{m}$, described in Equation \ref{eq:comm} ($j$ can be equal to $m$).

Considering $x_{ijkl}$ (resp., $y_{jkl}$) as the binary variable representing whether or not a client $c_i$ (resp., server $s$) executes on VM $vm_{jkl}$ of region $r_{jk}$ of provider $p_j$, the VMs execution and communication costs are respectively calculated by Equation \ref{eq:vm_costs} and Equation \ref{eq:transfer_costs} where  $cost_{jkl}$ is the fixed cost (in \$ per second) to use the VM $vm_{jkl}$ and $comm_{jm}$ is the cost to send and receive messages between provider $p_{j}$ and $p_{m}$. Note that as described in Equation \ref{eq:comm}, $j$ can be equal to $m$.
\begin{multline}
\resizebox{0.90\hsize}{!}{$vm\_costs = \sum_{c_i \in C}\sum_{p_j \in P}\sum_{r_{jk} \in R_{j}}\sum_{vm_{jkl} \in V_{jk}}(x_{ijkl} \times cost_{jkl}$} \\
\resizebox{0.90\hsize}{!}{$\times t_{m}) \  + \sum_{p_{j} \in P}{\sum_{r_{jk} \in R_{j}}\sum_{vm_{jkl} \in V_{jk}}(y_{jkl} \times cost_{jkl} \times t_m)}$} \label{eq:vm_costs}
\end{multline}
\begin{multline}
\resizebox{0.95\hsize}{!}{$comm\_costs = \sum_{c_i \in C}\sum_{p_{j} \in P}\sum_{r_{jk} \in R_{j}}\sum_{vm_{jkl} \in V_{jk}} \sum_{p_{m} \in P}\sum_{r_{mn} \in R_{m}}$} \\
\resizebox{0.55\hsize}{!}{$\sum_{vm_{mno} \in V_{mn}}(x_{ijkl} \times y_{mno} \times comm_{jm})$} \label{eq:transfer_costs}
\end{multline}
\begin{multline}
\resizebox{0.89\hsize}{!}{$comm_{jm} = (size(s\_msg_{train}) + size(s\_msg_{aggreg}))\times cost\_t_{m}$}\\
\resizebox{0.70\hsize}{!}{$ + (size(c\_msg_{train})\  + size(c\_msg_{test})) \times cost\_t_{j}$} \label{eq:comm}
\end{multline}

The monetary cost and the makespan in Equation \ref{eq:obj} can have different minimum and maximum values. 
Thus, we need to normalize both objectives to have the same range of 0-1. To this end, we use $T_{max}$ as the maximum possible makespan of an FL round for the current application in the current sets of providers, regions, and VMs as well as $cost_{max}$ as the maximum cost in an FL round, computed as Equation \ref{eq:cost_norm}. We obtain $cost_{max}$ by the sum of two following products: (i) the cost of hiring the most expensive VM times both $T_{max}$ and the number of tasks; and (ii) the most expensive communication costs between providers ($comm_{jm}$) times the number of clients.
%(i) the multiplication of the cost of hiring the most expensive VM by $T_{max}$ and by the number of tasks; and (ii) the multiplication of the most expensive communication costs between providers ($comm_{jm}$) by the number of clients.
%
\begin{multline}
cost_{max} = \max_{p_{j} \in P, r_{jk} \in R_{j}, vm_{jkl} \in V_{jk}} (cost_{jkl}) \times T_{max} \times (|C| + 1)\\
+ \max_{p_j, p_m \in P} (comm_{jm}) \times |C| \label{eq:cost_norm}
\end{multline}

%Our mathematical formulation has the following constraints. We ensure that our budget and deadline for an FL round are not violated through constraints \ref{r1} and \ref{r2}. Constraints \ref{r3} and \ref{r4} guarantee that each task executes on a single VM $vm_{jkl} \in V_{jk}$ in a region $r_{jk} \in R_{j}$ of a provider $p_{j} \in P$.

Our mathematical formulation must respect a set of constraints (\ref{r1} to \ref{r9}). 
Constraints \ref{r1} and \ref{r2} ensure that our budget and deadline for an FL round are not violated. Constraints \ref{r3} and \ref{r4} guarantee that each task executes on a single VM $vm_{jkl} \in V_{jk}$ in region $r_{jk} \in R_{j}$ of provider $p_{j} \in P$.
\begin{align}
& total\_costs \leq B_{round} \label{r1}
\end{align}
\begin{align}
& t_m \leq T_{round} \label{r2}
\end{align}
\begin{align}
& \resizebox{0.70\hsize}{!}{$\sum_{p_{j} \in P}{\sum_{r_{jk} \in R_{j}}{\sum_{vm_{jkl} \in V_{jk}} x_{ijkl}}} = 1 , \forall c_{i} \in C$} \label{r3}
\end{align}
\begin{align}
& \resizebox{0.60\hsize}{!}{$\sum_{p_{j} \in P}{\sum_{r_{jk} \in R_{j}}\sum_{vm_{jkl} \in V_{jk}} {y_{jkl}}} = 1$} \label{r4}
\end{align}

The next group of constraints (\ref{r5} to \ref{r8}) ensure that the solution does not use more GPUs and vCPUs per provider than %its limitations 
the maximum available bounds. Constraints \ref{r5} and \ref{r6} refer to the global provider bounds of available GPUs and vCPUs respectively while Constraints \ref{r7} and \ref{r8} refer to the same bounds but per region.
%the GPU and vCPU limits as well, but the regional ones.
% 
\begin{multline}
\resizebox{0.95\hsize}{!}{$\sum_{c_{i} \in C}{\sum_{r_{jk} \in R_j}{\sum_{vm_{jkl} \in V_{jk}} x_{ijkl} \times gpu_{jkl}}} + \sum_{r_{jk} \in R_j}\sum_{vm_{jkl} \in V_{jk}}(y_{jkl}$} \\
\resizebox{0.40\hsize}{!}{$\times gpu_{jkl}) \leq N\_GPU_j , \forall p_{j} \in P$} \label{r5}
\end{multline}
\begin{multline}
\resizebox{0.95\hsize}{!}{$\sum_{c_{i} \in C}{\sum_{r_{jk} \in R_j}{\sum_{vm_{jkl} \in V_{jk}} x_{ijkl} \times cpu_{jkl}}} +\sum_{r_{jk} \in R_j}\sum_{vm_{jkl} \in V_{jk}}(y_{jkl}$} \\
\resizebox{0.40\hsize}{!}{$\times cpu_{jkl}) \leq N\_CPU_j , \forall p_{j} \in P$} \label{r6}
\end{multline}
\begin{multline}
\resizebox{0.95\hsize}{!}{$\sum_{c_{i} \in C}{\sum_{vm_{jkl} \in V_{jk}}{x_{ijkl} \times gpu_{jkl}}} \ + \sum_{vm_{jkl} \in V_{jk}}{y_{jkl} \times gpu_{jkl}}$} \\
\resizebox{0.50\hsize}{!}{$\leq N\_L\_GPU_{jl} , \ \forall p_{j} \in P, \forall r_{jk} \in R_{j}$} \label{r7}
\end{multline}
\begin{multline}
\resizebox{0.95\hsize}{!}{$\sum_{c_{i} \in C}{\sum_{vm_{jkl} \in V_{jk}}{x_{ijkl} \times cpu_{jkl}}} \ + \sum_{vm_{jkl} \in V_{jk}}{y_{jkl} \times cpu_{jkl}}$} \\
\resizebox{0.50\hsize}{!}{$\leq N\_L\_CPU_{jl} , \ \forall p_{j} \in P, \forall r_{jk} \in R_{j}$} \label{r8}
\end{multline}

Constraint \ref{r9} ensures that $t_m$, the makespan of the FL round, will be at least equal to the total execution time of each client, which in turn is equal to the sum of the computational time $t\_exec_{ijkl}$ of the client in question, the communication time $t\_comm_{jklm}$ between this client's region and the server, and the aggregation time $t\_aggreg_{mno}$ of the server. The domain of the decision variables $x_{ijkl}$ and $y_{jkl}$ are determined in constraints \ref{r10} and \ref{r11}.
\begin{multline}
x_{ijkl} \times y_{mno} \times (t\_exec_{ijkl} + t\_comm_{jkmn} + t\_aggreg_{mno}) \leq t_{m}, \forall c_{i} \in C,\\ 
\forall p_{j},p_{m} \in P, \forall r_{jk} \in R_{j}, \forall vm_{jkl} \in V_{jk},
\forall r_{mn} \in R_{m},\  \forall vm_{mno} \in V_{mn} \label{r9}
\end{multline}
\begin{equation}
\resizebox{0.88\hsize}{!}{$x_{ijkl} \in \{0, 1\},\forall c_{i} \in C,\forall p_{j} \in P,\forall r_{jk} \in R_{j},\forall vm_{jkl} \in V_{jk}$} \label{r10}
\end{equation}
\begin{align}
\resizebox{0.70\hsize}{!}{$y_{jkl} \in \{0, 1\},\forall p_{j} \in P,\forall r_{jk} \in R_{j}, \forall vm_{jkl} \in V_{jk}$} & \label{r11}
\end{align}

% More details on the Initial Mapping module can be found in our conference paper~\cite{brum2022sbac}.

\subsection{Fault Tolerance module}

As previously explained, {\it Multi-FedLS} takes advantage of preemptible (or spot) VMs to reduce costs. The main drawback of these VMs is the revocation possibility at any time.
AWS usually sends a two-minute notification to the selected VM to notify about its revocation while Google Cloud, after only 30 seconds, terminates the VM. Thus, the application must handle possible VM revocations and 
internal errors during the execution, which lead to faulty FL tasks.
Note that the failure of a client has a different impact in the FL execution than the failure of the server. 

During the FL execution, the Fault Tolerance module monitors all tasks and handles eventual VMs revocations and runtime errors. When it detects any problem, it triggers the Dynamic Scheduler module (described in the next section) to select a new VM for the faulty task which was running in the revoked VM, informing if the task is the server or a client. 
After receiving from the Dynamic Scheduler the chosen VM to restart the task, the Fault Tolerance module launches it and continue to monitor all tasks within the FL application.
% When all tasks stop executing, the module stops monitoring the tasks and starts the VMs termination process.
When all tasks finish, the module stops monitoring the tasks and starts the VMs termination process.

Flower supports client failures during an FL round by making the server aggregate only the results received by the non-faulty clients. However, to start a new round, the FL server waits for a defined minimum number of available clients, which can be smaller than the total number of clients.
In Cross-Silo FL, there are usually few clients and not considering one of them at every round can have an impact in the learning outcomes. Thus, the FL server 
always waits for all clients to start the next round. 
% Note that t
There is no need to manually restart the model at the clients because in the beginning of each round, the server sends the current weights to all clients (Section \ref{sec:app_models}).

While Flower handles a client's error smoothly, 
%there is no fault tolerance implemented 
it does not tolerate the failure of the server. Furthermore, when it happens, all clients stop their execution after receiving an error. {\it Multi-FedLS} handles possible server faults, using fault-tolerant techniques on both server and client sides. 

Although Flower does not have an automatic way to save the model updates on the server side, it allows users to modify the server code to implement checkpointing during the server's aggregation phase\footnote{https://flower.dev/docs/saving-progress.html}. Our framework takes advantage of this feature by asking the server to checkpoint every $X$ round. 
Whenever a new checkpoint is stored on the server's VM local disk, it is transferred to another location asynchronously, which can be either a storage service or an extra VM.

On the client side, the aggregated weights received from the server in each round are stored in the VM's local disk, without sending it to another location. The clients' checkpoint is only used when the server's checkpoint is older than the clients' one.
Thus, to restart the server, it is necessary to verify if the server or the clients have the latest checkpoint. If it is the server checkpoint, the saved weights are sent to the new VM before restarting the server task and the FL server just reads the weights.
If it is the client one, the FL server restarts in the new VM and waits for any client to send its weights before starting the first round.

\subsection{Dynamic Scheduler module}
We denoted the faulty task as the task that was running in the revoked VM instance. The Dynamic Scheduler module adapts the objective function of the mathematical formulation of the Initial Mapping module and uses its value in a greedy heuristic to choose the new VM for the faulty task. It also takes advantage of the metrics computed by previous modules. 
Depending on the type of the faulty task, the Dynamic Scheduler computes differently the expected makespan and expected monetary costs, using a new instance to execute the faulty task. Algorithm \ref{alg:new-makespan} computes the new makespan using $t$ as the faulty task (which can be $s$ for the server or client $c_i$) and $vm_{jkl}$ as the instance to be allocated to task $t$. 
\begin{algorithm}
\caption{\small{Makespan Re-calculation}} \label{alg:new-makespan}
\begin{algorithmic}
\scriptsize
\renewcommand{\algorithmicrequire}{\textbf{Input:}}
\renewcommand{\algorithmicensure}{\textbf{Output:}}
\REQUIRE $t, vm_{jkl}$
\STATE $max\_makespan \gets -\infty$
\IF{$t = s$ }\COMMENT{Faulty task is the server} \label{alg:new-makespan-task-type}
    \FOR{$c_i \in C$} \label{alg:new-makespan-loop-server-init}
        \STATE $current\_vm_{mno} \gets current\_map_{c_i}$
        \STATE $total\_time \gets t\_exec_{imno} + t\_comm_{mnjk} + t\_aggreg_{jkl}$
        \IF{$total\_time > max\_makespan$}
            \STATE $max\_makespan \gets total\_time$
        \ENDIF
    \ENDFOR \label{alg:new-makespan-loop-server-end}
\ELSE{ }\COMMENT{Faulty task is client $c_t$} \label{alg:new-makespan-client}
    \STATE $current\_server\_vm_{mno} \gets current\_map_{s}$
    \STATE $max\_makespan \gets t\_exec_{tjkl} + t\_comm_{jkmn} + t\_aggreg_{mno}$
    \FOR{$c_i \in C\backslash \{c_t\}$} \label{alg:new-makespan-loop-client-init}
        \STATE $current\_vm_{pqr} \gets current\_map_{c_i}$ \label{alg:new-makespan-client-server-vm}
        \STATE $total\_time \gets t\_exec_{ipqr} + t\_comm_{pqmn} + t\_aggreg_{mno}$
        \IF{$total\_time > max\_makespan$}
            \STATE $max\_makespan \gets total\_time$
        \ENDIF
    \ENDFOR \label{alg:new-makespan-loop-client-end}
\ENDIF
\STATE \textbf{return} $max\_makespan$
\end{algorithmic}
\end{algorithm}

The FL round makespan is defined by the client which takes longer to send its weights to the server. The delay is related to a longer training time and/or communication time. Thus, the makespan is obtained by computing the higher time a client needs to train and communicate with the server along with the server's aggregation time. 
If the faulty task is the server, then $vm_{jkl}$ is the new server instance and  the loop swaps through all clients $c_i \in C$ to compute the total expected execution time of each client and store the highest one as the new makespan. If the faulty task is a client, 
the current server instance is available in variable $current\_server\_vm_{mno}$ and the makespan of the faulty task is computed by using $vm_{jkl}$ as the client's instance. Then, the loop swaps through all clients $c_i \in C$, excluding the current task to check if any of the current clients have a higher makespan. 

Algorithm \ref{alg:new-cost} shows how to compute the new cost, which receives the same inputs as the previous algorithm, i.e., the faulty task $t$ and the new instance $vm_{jkl}$ as well as the makespan in $ms$.
\begin{algorithm}
 \caption{\small{Financial Cost Re-calculation}} \label{alg:new-cost}
\begin{algorithmic}
\scriptsize
\renewcommand{\algorithmicrequire}{\textbf{Input:}}
\renewcommand{\algorithmicensure}{\textbf{Output:}}
\REQUIRE $t, vm_{jkl}, makespan$
\STATE $total\_cost \gets 0.0$
\IF{$t = s$ }\COMMENT{Faulty task is the server}
    \STATE $total\_cost \gets total\_cost + cost_{jkl} \times ms$
    \FOR{$c_i \in C$}
        \STATE $current\_vm_{mno} \gets current\_map_{c_i}$
        \STATE $total\_cost \gets total\_cost + cost_{mno} \times ms + comm_{jm}$ \COMMENT{$comm_{jm}$ as Equation \ref{eq:comm}}
    \ENDFOR
\ELSE{ }\COMMENT{Faulty task is client $c_t$}
    \STATE $current\_server\_vm_{mno} \gets current\_map_{s}$ \label{alg:new-cost-server-client}
    \STATE $total\_cost \gets total\_cost + cost_{jkl} \times ms + comm_{mj}$ \COMMENT{$comm_{mi}$ as Equation \ref{eq:comm}}
    \FOR{$c_i \in C\backslash \{c_t\}$}
        \STATE $current\_vm_{pqr} \gets current\_map_{c_i}$
        \STATE $total\_cost \gets total\_cost + cost_{pqr} \times ms + comm_{mp}$ \COMMENT{$comm_{mp}$ as Equation \ref{eq:comm}}
    \ENDFOR
\ENDIF
\STATE \textbf{return} $total\_cost$
\end{algorithmic}
\end{algorithm}

The total cost in a single FL round comprises the execution cost of each task and message exchange costs of all clients to the server. Thus, in Algorithm \ref{alg:new-cost} the execution cost of the server is firstly computed and then the execution and message exchange costs of all clients are also computed. The execution cost of an instance is the multiplication of the makespan and the instance cost while the message exchange cost is computed using $comm_{jm}$ (Equation \ref{eq:comm}), the cost to exchange the FL messages between the server's and the client's providers. Equally to the previous algorithm, when the faulty task $t$ is the server, the server provider $p_j$ is obtained from the input instance $vm_{jkl}$. Otherwise, if the faulty task is a client, the server's provider $p_m$ is obtained from the current server instance.

Algorithm \ref{alg:greedy-ds} implements a greedy heuristic for choosing a new instance for a server or a client after a revocation, receiving as input $t$, the faulty task, $I_{t}$, the current set of all possible instances for task $t$ and $old\_vm_{jkl}$, the current revoked VM instance. Note that if the faulty task is the server then $t = s$ and $I_{t} = I_{s}$, the set of possible server instances; otherwise, the task
\begin{algorithm}
 \caption{\small{Instance Selection}} \label{alg:greedy-ds}
\begin{algorithmic}
\scriptsize
\renewcommand{\algorithmicrequire}{\textbf{Input:}}
\renewcommand{\algorithmicensure}{\textbf{Output:}}
\REQUIRE $t, I_{t}, old\_vm_{jkl}$
\STATE $remove\_vm\_from\_set(I_{t}, old\_vm_{jkl})$
\STATE $min\_value \gets \infty$
\FOR{\textbf{each} $vm_{mno} \in I_{t}$} \label{alg:loop-init-greedy}
    \STATE $makespan \gets computes\_new\_makespan(t, vm_{mno})$ \COMMENT{Algorithm \ref{alg:new-makespan}}
    \STATE $cost \gets compute\_expected\_cost(makespan, t, vm_{mno})$ \COMMENT{Algorithm \ref{alg:new-cost}}
    \STATE $value \gets \alpha \times (cost / cost_{max}) +  (1 - \alpha) \times (makespan / T_{max})$
    \IF {$value < min\_value$}
        \STATE $new\_instance \gets vm_{mno}$
        \STATE $min\_value \gets value$
    \ENDIF
\ENDFOR \label{alg:loop-end-greedy}
\STATE \textbf{return} $new\_instance$
\end{algorithmic}
\end{algorithm}
 is client $c_i$, then $t = c_i$ and $I_{t} = I_{c_i}$, the set of possible instances for client $c_i$. Moreover, in the first execution of the Dynamic Scheduler, $I_{t}$ is the sum of all $V_{jk}$ sets for all regions $r_{jk} \in R_j$ for all providers $p_j \in P$.

The first step in Algorithm \ref{alg:greedy-ds} is to remove the revoked VM from the set of possible VMs. From previous experiments, we observed that once the instance type is revoked in a region in AWS, it cannot be reallocated in the same region immediately~\cite{brum2021fault}. Thus, we assume this behavior to all cloud providers and remove the $old\_vm_{jkl}$ from $I_{t}$. Note that we do not use the same set for all tasks, or not even for all clients, as they can have different expected execution times which can correspond to different selections in both Initial Mapping and Dynamic Scheduler modules. After removing the revoked VM, our algorithm starts the loop iterating on all VMs from $I_{t}$ to select the VM to restart task $t$. The makespan and cost are computed with Algorithms \ref{alg:new-makespan} and \ref{alg:new-cost}, respectively, using $vm_{mno}$ as the instance for the task $t$. After that, the sum of the makespan and cost is done by using the same weight as the objective function of the Initial Mapping module (Equation \ref{eq:obj}), $\alpha$ for the normalized cost and $1-\alpha$ for the normalized makespan.

\section{Experimental environment and results} \label{sec:experiments}

This section describes the experimental environment using three different applications and the obtained results.

\subsection{Applications}

We selected three applications to execute in our framework: a real-world use-case application and two applications from an FL benchmark. The use-case application, detailed in~\cite{brum2021wcc,brum2021eradrj}, identifies Tumor-Infiltrating Lymphocytes (TILs) in high-resolution human tissue images and generates maps depicting their spatial distribution. However, the high pixel count (up to 100K$\times$100K) of Whole Slide Tissue Images (WSIs) makes it challenging to extract information from the thousands of WSIs available. Saltz~\etal~\cite{saltz:2018} developed a deep learning application  for TIL classification and analysis that uses the VGG16 model, a convolutional neural network (CNN) model used for image recognition. We evaluate %our proposal 
{\it Multi-FedLS} by focusing on the CNN training step, assuming that the WSIs have already been classified by experts and divided into patches. The TIL application has 4 clients with 948 training samples and 522 test samples each.

The other two applications are from LEAF~\cite{caldas2019leaf}, a hub of benchmarks for different FL settings, mostly focused on Cross-Device FL (thousands of clients with few samples). We selected two datasets, the Shakespeare and the FEMNIST ones, and we adapted them to a Cross-Silo FL setting (fewer clients with more samples). 

The Shakespeare dataset is based on the book \textit{The Complete Works of William Shakespeare}, with each character of each play being a different client. We selected clients with more than 18000 samples combining the training and test datasets, which left us with 8 clients. Their training datasets vary from 16488 to 26282 samples and their test datasets vary from 1833 to 2921 samples. This dataset is used to predict the next character in the sentence, and we implemented the reference model in LEAF~\cite{caldas2019leaf}, which uses an embedding of dimension 8 and a Long Short Term Memory (LSTM) network of two layers containing 256 units each. This LSTM network is a recurrent neural network that saves information from the previous sample to be used by the current sample. We chose the Shakespeare application because the clients have bigger datasets with a small model to train.
        
The FEMNIST dataset creates a federated scenario to the extended MNIST dataset, containing both handwritten letters and digits of size 28$\times$28. In this dataset, each client has the handwritten characters of a single user. We selected users with more than 440 samples in total, which left us with 5 clients. After that, we replicated each client's datasets to double their number of samples. Thus, the training datasets of these five clients vary from 796 to 1050 samples and their test datasets vary from 90 to 118 samples. To predict the character in an image in the FEMNIST dataset, we create a more robust CNN than the reference one in LEAF~\cite{caldas2019leaf} using 2 convolutional layers followed by 10 fully connected layers with 4096 neurons each. We chose this application because the clients have smaller datasets with a more robust model to train compared to the Shakespeare application. Both applications allow us to understand the behaviour of our {\it Multi-FedLS} with a wide range of different characteristics (small vs. big dataset; simple vs. complex model).% that can impact the execution time and costs.

All three applications used FedAvg as the server aggregation method.

\subsection{Experimental setup}
\label{sec:expsetup}

In our previous work~\cite{brum2022sbac, brum2022wcc}, experiments were carried out on two commercial clouds: Amazon Web Services (AWS) and Google Cloud Provider (GCP). Both of them %do not increase 
restrict our GPU's quotas, providing only 4 simultaneous GPUs. Thus, for scalability sake, most experiments were conducted on CloudLab~\cite{cloudlab_paper}, a platform that allows the simulation of cloud environments. It behaves similarly to a real cloud platform, with predefined instances types in clusters in at least three different US states\footnote{https://www.cloudlab.us/hardware.php}. 
% However, CloudLab instantiates the resource isolating its hardware from other users' resources. 
However, contrary to real providers, CloudLab does not use virtualization but a \textit{bare metal} approach to isolate the instances requested by a user from others.
%This \textit{bare metal} approach reduces the performance variation of real cloud providers described by Ward and Barker~\cite{Ward2014} and found in our previous results~\cite{brum2022sbac}. 
Moreover, 
CloudLab does not limit the number of vCPUs and GPUs per region allocated by users. On the other hand, they use a reservation system, in which users must specify which instances they want to use, and the latter will be available during all the scheduled reservation.
% there are no user limits regarding the number of vCPUs and GPUs per region in CloudLab, but users need to request a reservation to a specific instance type so it will be available for them during all  experiments.
CloudLab has five different clusters, with two of them in the same US state. We simulate two different clouds by grouping the clusters into two clouds: \textit{Cloud A} and \textit{Cloud B} with three and two clusters respectively. Each cluster simulates a different region inside the respective cloud.

%Some machine types in CloudLab have a huge amount of vCPUs and memory being too expensive to execute the server. They are also not a proper choice for clients  due to the lack of GPUs running them extremely slow. Thus, we subsampled 13 CloudLab instances.  
% Remove after gihub - \footenote{See xxx for all suplementary material.}

Table \ref{tab:setup} summarizes the instance selection of the testbed.
The VMs that have GPUs are $vm_{126}$ with a P100 (12 GB of memory) and $vm_{138}$ with a Tesla V100S (32 GB of memory).
As CloudLab does not charge for the use of its instances, we computed the on-demand price for each VM based on Google Cloud Provider's policy. 
\begin{table}[h!tpp]
\caption{Instance types selected} \label{tab:setup}
\renewcommand{\arraystretch}{0.95}
\scriptsize
\begin{center}
\begin{tabular}{|c|c|c|c|c|c|c|c|}
\hline
\multirow{2}{*}{Cloud} & \multirow{2}{*}{Cluster/Region} & \multirow{2}{*}{VM} & \multirow{2}{*}{vCPUS} & \multirow{2}{*}{\begin{tabular}[c]{@{}c@{}}RAM\\(GB)\end{tabular}} & \multicolumn{2}{c|}{\begin{tabular}[c]{@{}c@{}}Costs per hour (\$)\end{tabular}} & \multirow{2}{*}{ID} \\ 
\cline{6-7}
 & & & & & On-demand & Spot & \\

% Cloud & Cluster/Region & VM & vCPUS & RAM (GB) & On-demand (\$) & Spot (\$) & ID \\
\hline
\multirow{9}{*}{Cloud A} & \multirow{3}{*}{Utah}           & 
                          c6525-25g     &  32 & 128 & 1.670 & 0.501 & $vm_{112}$ \\
\cline{3-7}
                          &                                 & m510          &  16 &  64 & 0.835 & 0.250 & $vm_{114}$ \\
\cline{3-7}
                          &                                 & xl170         &  20 &  64 & 0.971 & 0.291 & $vm_{115}$ \\
\cline{2-7}
                          & \multirow{4}{*}{Wisconsin}      & c220g1        &  32 & 128 & 1.670 & 0.501 & $vm_{121}$ \\
\cline{3-7}
                          &                                 & c220g2        &  40 & 160 & 2.087 & 0.626 & $vm_{122}$ \\
\cline{3-7}
                          &                                 & c240g1        &  32 & 128 & 1.670 & 0.501 & $vm_{124}$ \\
\cline{3-7}
                          &                                 & c240g5        &  40 & 192 & 4.693 & 1.408 & $vm_{126}$ \\
\cline{2-7}
                          & \multirow{2}{*}{Clemson}              & dss7500       &  24 & 128 & 1.398 & 0.419 & $vm_{135}$ \\
\cline{3-7}
                          &                                 & r7525         & 128 & 512 & 11.159 & 3.348 & $vm_{138}$ \\
\hline
\multirow{4}{*}{Cloud B}  & \multirow{2}{*}{APT}            & c6220         &  32 &  64 & 1.283 & 0.385 & $vm_{211}$ \\
\cline{3-7}
                          &                                 & r320          &  12 &  16 & 0.574 & 0.172 & $vm_{212}$ \\
\cline{2-7}
                          & \multirow{2}{*}{Massachusetts}  & rs440         &  64 & 192 & 2.837 & 0.851 & $vm_{221}$ \\
\cline{3-7}
                          &                                 & rs630         &  40 & 256 & 2.349 & 0.705 & $vm_{222}$ \\
\hline
\end{tabular}
\end{center}
\end{table}
We used the values of December 2022 of the computer-optimized instances of GCP to Cloud A, \$0.03398 per vCPU and \$0.00455 per GB of RAM per hour, and general-purpose instances to Cloud B, \$0.031611 per vCPU and \$0.004237 per GB of RAM per hour. 
% GPU's architectures have different prices per hour per GPU being \$1.46 for P100, and \$2.48 for V100.
GPU's architectures have different prices per hour being \$1.46 per P100, and \$2.48 per V100.
The spot price of a VM of a given type was set by considering a 70\% discount on the price of the on-demand VM of the same type. 

% \vspace{1cm}
\subsection{Pre-Scheduling slowdowns}
We computed the computational slowdown for all instances in Table \ref{tab:setup} using $vm_{121}$ as the baseline VM. One client of the TIL use-case application executed 38 training and 21 test samples. Table \ref{tab:slowdown_vms} shows the obtained slowdowns. 
Note that all VMs have different execution times, varying the slowdown from 0.045 and 2.328.
\begin{table}[h!tbp]
\caption{Time of one client with five local epochs,  dataset stored in Utah region of Cloud\_A} \label{tab:slowdown_vms}
\scriptsize
\begin{center}
\begin{tabular}{|c|c|l|r|r|r|r|c|}
\hline
\multirow{2}{*}{Cloud} & \multirow{2}{*}{Region} & \multirow{2}{*}{VM ID} & \multicolumn{2}{c|}{Training time} & \multicolumn{2}{c|}{Test time} & \multirow{2}{*}{Slowdown} \\
\cline{4-7}
 & & & 1º r. & 2º r. & 1º r. & 2º r. & \\
\hline
\multirow{9}{*}{Cloud A} & \multirow{3}{*}{Utah}       & $vm_{112}$ & 123.12 & 120.93 &  1.61 &  1.47 & 1.064 \\
\cline{3-8}
                          &                                 & $vm_{114}$ & 163.16 & 158.95 &  4.71 &  4.62 & 1.422 \\
\cline{3-8}
                          &                                 & $vm_{115}$ & 113.22 & 110.32 &  2.95 &  2.86 & 0.984 \\
\cline{2-8}
                          & \multirow{4}{*}{Wisconsin}      & $vm_{121}$ & 119.89 & 112.83 &  2.30 &  2.22 & 1.000 \\
\cline{3-8}
                          &                                 & $vm_{122}$ & 139.04 & 131.74 &  1.93 &  1.96 & 1.162 \\
\cline{3-8}
                          &                                 & $vm_{124}$ & 119.05 & 110.45 &  2.23 &  2.12 & 0.970 \\
\cline{3-8}
                          &                                 & $vm_{126}$ &  16.37 &   4.53 &  1.44 &  0.62 & 0.045 \\
\cline{2-8}
                          & \multirow{2}{*}{Clemson}      & $vm_{135}$ & 128.46 & 122.39 &  2.79 &  2.67 & 1.087 \\
\cline{3-8}
                          &                                 & $vm_{138}$ &  71.67 &  60.14 &  5.39 &  5.24 & 0.568 \\
\hline
\multirow{4}{*}{Cloud B}  & \multirow{2}{*}{APT}            & $vm_{211}$ & 147.79 & 141.62 &  4.22 &  4.26 & 1.268 \\
\cline{3-8}
                          &                                 & $vm_{212}$ & 263.89 & 256.73 & 11.18 & 11.13 & 2.328 \\
\cline{2-8}
                          & \multirow{2}{*}{Massachusetts}  & $vm_{221}$ &  94.23 &  92.42 &  1.26 &  1.20 & 0.814 \\
\cline{3-8}
                          &                                 & $vm_{222}$ & 112.44 & 103.59 &  1.91 &  1.75 & 0.916 \\
\hline
\end{tabular}
\end{center}
\end{table}

We also computed the communication slowdowns in Table \ref{tab:slowdown_comm} using the pair of regions Cloud\_B\_APT-Cloud\_B\_APT as the baseline. The training and test phases exchange a total of 2GB in messages and a little more than 1GB respectively. Note that communication times vary a lot in CloudLab, with the slowdown ranging from 0.372 to 24.731.

\begin{table}[h!tbp]
\caption{Time between each pair of regions} \label{tab:slowdown_comm}
\scriptsize
\begin{center}
\begin{tabular}{|c|r|r|r|}
\hline
\multirow{2}{*}{Pair of regions} & \multicolumn{2}{c|}{Comm. times (s)} & \multirow{2}{*}{Slowndown} \\
\cline{2-3} 
                & Training  & Test  &        \\
\hline
Cloud\_B\_APT-Cloud\_B\_APT         &      5.61 &  3.05 &  1.000 \\
\hline
Cloud\_B\_APT-Cloud\_A\_Clemson     &     12.05 &  5.94 &  2.078 \\
\hline
Cloud\_B\_APT-Cloud\_B\_Mass        &    106.90 & 54.51 & 18.641 \\
\hline
Cloud\_B\_APT-Cloud\_A\_Utah        &      4.84 &  2.58 &  0.857 \\
\hline
Cloud\_B\_APT-Cloud\_A\_Wis         &     16.19 &  7.64 &  2.752 \\
\hline
Cloud\_A\_Clemson-Cloud\_A\_Clemson &      5.36 &  2.91 &  0.954 \\
\hline
Cloud\_A\_Clemson-Cloud\_B\_Mass    &     75.63 & 32.31 & 12.464 \\
\hline
Cloud\_A\_Clemson-Cloud\_A\_Utah    &     11.39 &  5.34 &  1.932 \\
\hline
Cloud\_A\_Clemson-Cloud\_A\_Wis     &      6.65 &  3.53 &  1.175 \\
\hline
Cloud\_B\_Mass-Cloud\_B\_Mass       &      5.23 &  2.81 &  0.929 \\
\hline
Cloud\_B\_Mass-Cloud\_A\_Utah       &     86.08 & 35.95 & 14.092 \\
\hline
Cloud\_B\_Mass-Cloud\_A\_Wis        &    138.31 & 75.85 & 24.731 \\
\hline
Cloud\_A\_Utah-Cloud\_A\_Utah       &      2.07 &  1.15 &  0.372 \\
\hline
Cloud\_A\_Utah-Cloud\_A\_Wis        &     21.81 & 10.57 &  3.738 \\
\hline
Cloud\_A\_Wis-Cloud\_A\_Wis         &      5.77 &  3.08 &  1.022 \\
\hline
\end{tabular}
\end{center}
\end{table}

\subsection{Validation of CloudLab environment}

We validate the CloudLab environment using the most robust application, the real-world TIL, in order to compare the obtained results with the ones of our previous work of \cite{brum2022sbac}. The baseline execution time (training and test phases) for each client is 2765.4 seconds and the communication baseline is 8.66 seconds. %We execute the 
TIL application executes for 10 communication rounds since such number of rounds has presented the best ML metrics~\cite{brum2021eradrj}. We assume the transfer costs inside both clouds are the same as the ones evaluated for Google Cloud Provider, \$0.012 per sent GB. 

The Initial Mapping module shows that the optimized configuration to run the TIL application in CloudLab is composed of a VM $vm_{121}$ for the server and four VMs $vm_{126}$ for clients. It also yields that the runtime would be 22:38 minutes and the costs would be \$15.44. The average runtime value for three executions 
% in CloudLab 
is 24:47 minutes and the average cost is \$16.18.

% Table \ref{tab:initial_mapping_validation} shows the model output for the FL runtime and costs along with the values for the real execution. We emphasize that the values shown are the average of three executions. This experiment assumes that all VMs are on-demand, and thus, we do not use any Fault Tolerance mechanism.

% \begin{table}[h!tpp]
% \caption{Validating CloudLab with Initial Mapping module using the TIL application} \label{tab:initial_mapping_validation}
% \scriptsize
% \begin{center}
% \begin{tabular}{|l|r|r|r|r|}
% \hline
% \multicolumn{1}{|c|}{\multirow{2}{*}{Setup}} & \multicolumn{2}{c|}{Model output} & \multicolumn{2}{c|}{Real Execution} \\
% \cline{2-5} 
% \multicolumn{1}{|c|}{}                       & \multicolumn{1}{c|}{Runtime}      & \multicolumn{1}{c|}{Costs} & \multicolumn{1}{c|}{Runtime} & \multicolumn{1}{c|}{Costs} \\
% \hline
% \begin{tabular}[c]{@{}l@{}}server in $vm_{121}$ and\\ clients in $vm_{126}$\end{tabular} & 0:22:38 & \$15.44 & 0:24:47 & \$16.18 \\ 
% \hline
% \end{tabular}
% \end{center}
% \end{table}

We observed that the model has a similar execution time and cost compared to the real one for the TIL use-case application. The difference is 8.69\% in the execution time and 4.53\% in costs, validating this new test environment with the same setup used in~\cite{brum2022sbac}. 

However, we also noticed that our framework took longer to start the FL execution in all VMs compared to real cloud providers. We compared the time to execute the TIL application in CloudLab, in AWS, and in GCP, using the results presented in~\cite{brum2022sbac}. The FL execution time in AWS corresponds to 91.45\% of the total {\it Multi-FedLS} execution time when there is no revocation. In CloudLab it is only 31.67\%, which can be explained by the \textit{bare-metal} approach, which increases the VM preparation time (2:34 in AWS versus 39:43 in CloudLab). The same behavior is found when compared to GCP. There is a huge difference in the preparation time (13:35 in GCP versus 39:43 in CloudLab). In GCP, the FL execution time corresponds to 53.30\% of the {\it Multi-FedLS} execution time. Besides, one disadvantage of CloudLab is that once the instance is terminated, the data modified in it gets lost, and we need to download the results before terminating the VMs, which adds more than 20 minutes in the {\it Multi-FedLS} execution. In AWS and GCP, we the produced data were stored in an extra volume not deleted with the VM. It was thus possible to download them afterwards. 

\subsection{Checkpoint interval overhead}

\begin{figure}[h!tpp]
 \centering
 \includegraphics[width=0.70\textwidth]{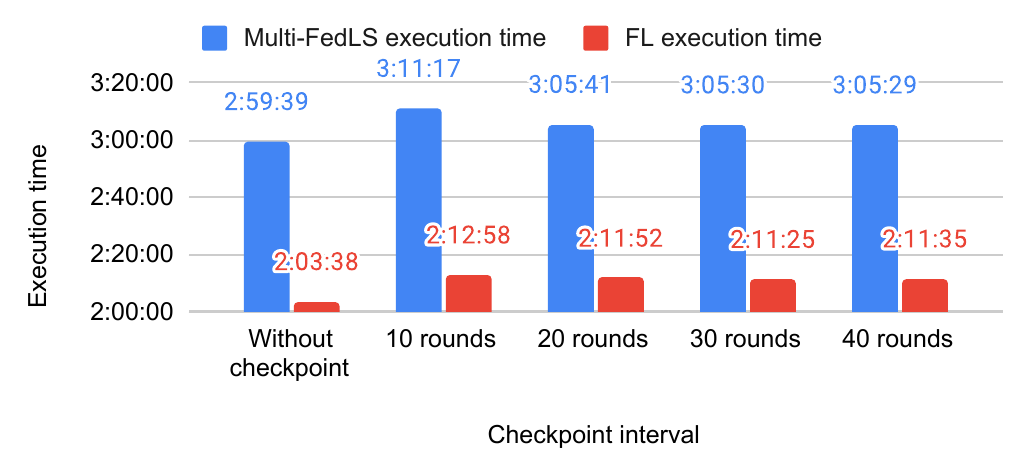}
 \caption{Server checkpoint overhead.}
 \label{fig:server_ckpt}
\end{figure}

The next experiment evaluates the impact of the Fault Tolerance module on the FL execution time and the total framework time using the TIL use-case application because, among all applications, it has the biggest model and the most costly checkpoint (504 MB). %We increase the number of rounds to reflect on the execution time and make the use of the Fault Tolerance and Dynamic Scheduler modules more reasonable. 
The number of rounds of the application was increased aiming a longer execution time as well as the calling of the Fault Tolerance and Dynamic Scheduler modules with different frequencies. 
We distinguish two 
%different 
types of checkpoints: one on the server every $X$ rounds and the client checkpoint every round. We evaluate them separately. Figure \ref{fig:server_ckpt} shows the average of {\it Multi-FedLS} and the FL execution time of three executions varying the $X$ within 10, 20, 30, and 40 rounds. 

We observe that the frequency of checkpointing impacts both execution times similarly. The overhead compared to the FL execution without checkpoint varies from 6.29\% (30 rounds) to 7.55\% (10 rounds).
This overhead is mostly due to the saving time to the local disk, as the checkpoints sending to another location overlaps the server's waiting for clients' messages.

Since clients do not send their checkpoints to an external location, we compute the clients' checkpoint overhead saving the weights in the disk after every evaluation step. The average execution time from three runs was 3:03:44 to the {\it Multi-FedLS} time and 2:06:20 when considering only the FL execution time, which corresponds to a 2.17\% overhead compared to no checkpoint execution.

\subsection{Failure simulation}

Failure simulation experiments were conducted with the three applications in order to observe the behaviour of Multi-FedLS as a whole at failure presence. %We first present the results for the real-world application and then the results for the benchmark applications.

\subsubsection{TIL application}

To validate the Dynamic Scheduler module, we simulated the VM revocation using a Poisson distribution~\cite{ahrens1974poisson} with a revocation rate $\lambda = 1/k_r$, where $k_r$ is the average time between failures in seconds. We observed patterns in the revocation frequency in instances with GPU in AWS~\cite{brum2021fault} which gave us two different values to our experimental $k_r$, 2 hours and 4 hours. Thus, there are two revocation rates: (i) $k_r = 7200$ and  $\lambda = 1/7200$, and (ii) $k_r = 14400$ and $\lambda = 1/14400$.

The revocation of a client and the server impact differently in the FL execution. Thus, we considered two simulation scenarios to explore such a difference. The first scenario (server and clients on spot VMs) uses preemptible/spot VMs for all tasks. The second scenario (server on an on-demand VM and clients on spot VMs) uses an on-demand VM to the server, which increases the costs compared to the first scenario, but ensures the reliability of the server task. 
%Another scenario could be the opposite with clients using on-demand VMs and only the server in a spot VM. But we do not consider this scenario as it has higher costs compared to the other scenarios and low reliability.

Table \ref{tab:failure_simulation_diff_instances} shows the results for three executions for the two scenarios. %where the columns are the simulation scenario (Scenario), the termination rate in question, the average number of revocations (Avg \# revoc.), the average execution time (Avg exec. time), and the average total costs (Avg total costs). 
\begin{table}[h!tpp]
\caption{Failure simulation using TIL application changing to another VM instance in CloudLab} \label{tab:failure_simulation_diff_instances}
\scriptsize
\begin{center}
\begin{tabular}{|c|c|r|r|r|}
\hline
Scenario                      & \begin{tabular}[c]{@{}c@{}}Termination\\ rate\end{tabular} & \multicolumn{1}{|c|}{\begin{tabular}[c]{@{}c@{}}Avg \#\\ revoc.\end{tabular}} & \multicolumn{1}{c|}{\begin{tabular}[c]{@{}c@{}}Avg exec.\\ time\end{tabular}} & \multicolumn{1}{c|}{\begin{tabular}[c]{@{}c@{}}Avg total\\ costs\end{tabular}} \\
\hline
\multirow{2}{*}{Server and clients on spot VMs} & $k_r = 7200$ & 3.67 & 10:01:46 & \$81.12 \\
\cline{2-5}
                              & $k_r = 14400$ & 0.00 &  3:04:37 & \$15.64 \\
\hline
\multirow{2}{*}{\begin{tabular}[c]{@{}c@{}}Server on an on-demand VM\\and clients on spot VMs\end{tabular}} & $k_r = 7200$ & 1.00 &  6:31:44 & \$55.60 \\
\cline{2-5}
                              & $k_r = 14400$ & 0.00 &  3:05:39 & \$19.27 \\
\hline
%\multirow{2}{*}{\begin{tabular}[c]{@{}c@{}}Server on a spot VM\\and clients on on-demand VMs\end{tabular}}  & $k_r = 7200$ & 0.00 &  3:00:44 & \$48.11 \\
%\cline{2-5}
%                              & $k_r = 14400$ & 0.00 &  3:04:16 & \$49.13 \\
%\hline
\end{tabular}
\end{center}
\end{table}
As a comparison, if all tasks execute on on-demand VMs without checkpoints, the execution time is 2:59:39 and the total costs are \$50.51.

Our dynamic scheduler always chooses the same instance types when there was a revocation. Clients start on a VM $vm_{126}$ and restart on a VM $vm_{138}$. The server starts on a VM $vm_{121}$ and restarts in a VM $vm_{212}$. There was at most one revocation per task in each execution.
In the first scenario, %where all tasks execute in spot VMs, 
two of the three executions had four revocations (3 clients and the server) and the last one had three revocations (2 clients and the server), having an average revocation number of 3.67. 
In the second scenario, %where the server executes in an on-demand VM and the clients are in spot VMs,
one of the executions had two revocations, another one revocation and the last one did not have any revocation, resulting in one average revocation.

We can observe that spot VMs is not an advantage in CloudLab, 
since, given any two VMs, client execution times are not similar.
%as no two VMs execute the clients at similar times. 
The $vm_{138}$ and $vm_{126}$ have slowdown of 0.568 and 0.045, respectively (Table \ref{tab:slowdown_vms}). 
Moreover, the time to instantiate the VMs also impacts the execution time, and, eventually, the total costs with spot VMs surpass the costs when using only on-demand VMs.

Real commercial cloud providers usually have VMs with similar or even equivalent configurations in different cloud regions. For example, AWS provides \textit{g4dn.2xlarge} instances (that have Turing GPUs) in 23 of its regions\footnote{https://aws.amazon.com/ec2/pricing/on-demand/, accessed in May 2023} and GCP provides Turing GPUs in seven of its regions\footnote{https://cloud.google.com/compute/docs/gpus/gpu-regions-zones, acessed May 2023}. These equivalent machines tend to have similar execution times (with a slight difference), as observed in the slowdowns in our previous paper~\cite{brum2022sbac}. On CloudLab, each instance type has completely different hardware configurations, but it is possible to use the same instance type immediately after a revocation.
Our dynamic scheduler does not remove the VM that was revoked from the available instance types (first line of Algorithm \ref{alg:greedy-ds}), %and it allows to choose 
allowing, thus, the selection of the same instance type in every revocation.

Table \ref{tab:failure_simulation_same_instances} presents the results from three executions restarting in the same instance type.%, where columns have the same meaning as Table \ref{tab:failure_simulation_diff_instances}. 

\begin{table}[h!tpp]
\caption{Failure simulation using TIL application changing to the same VM instance in CloudLab} \label{tab:failure_simulation_same_instances}
\scriptsize
\begin{center}
\begin{tabular}{|c|c|r|r|r|}
\hline
Scenario                      & \begin{tabular}[c]{@{}c@{}}Termination\\ rate\end{tabular} & \multicolumn{1}{c|}{\begin{tabular}[c]{@{}c@{}}Avg \#\\ revoc.\end{tabular}} & \multicolumn{1}{c|}{\begin{tabular}[c]{@{}c@{}}Avg exec.\\ time\end{tabular}} & \multicolumn{1}{c|}{\begin{tabular}[c]{@{}c@{}}Avg total\\ costs\end{tabular}} \\
\hline
\multirow{2}{*}{Server and clients on spot VMs} & $k_r = 7200$ & 1.33 & 4:14:16 & \$22.55 \\
\cline{2-5}
                              & $k_r = 14400$ & 0.00 & 3:04:35 & \$15.64 \\
\hline
\multirow{2}{*}{\begin{tabular}[c]{@{}c@{}}Server on an on-demand VM\\and clients on spot VMs\end{tabular}} & $k_r = 7200$ & 0.33 & 3:14:38 & \$20.16 \\
\cline{2-5}
                              & $k_r = 14400$ & 0.00 & 3:01:49 & \$18.99 \\
\hline
%\multirow{2}{*}{\begin{tabular}[c]{@{}c@{}}Server on a spot VM\\and clients on on-demand VMs\end{tabular}}  & $k_r = 7200$ & 0.33 & 5:33:27 & \$95.73 \\
%\cline{2-5}
%                              & $k_r = 14400$ & 0.00 & 3:12:57 & \$49.28 \\
%\hline
\end{tabular}
\end{center}
\end{table}

In the first scenario with the termination rate of 1$/$(2 hours) ($1/7200$), the first, second, and third executions had two clients, one client,  and one server revocations respectively,
%one execution had two clients' revocations, another just one client revocation and the third one the server revocation,
totaling an average of 1.33 revocations. 
In the second scenario with the same termination rate, just one execution revoked one client, having 0.33 as the average revocation number.

From these results, we can observe that the revocation of a client impacts less the execution time than the server's revocation (3:14:38 vs. 4:14:16). 
Regarding the server execution in an on-demand VM vs. all tasks in spot VMs, we can see that, when there is no revocation, costs increase by 21.30\% (\$18.99 vs. \$15.64), %while when there are revocations, 
otherwise they decrease by 10.58\% (\$20.16 vs. \$22.55).

\subsubsection{Benchmarks}

The Shakespeare (resp., FEMINIST) applications were executed with 20 (resp., 100) rounds and 20 (resp., 100) epochs per round, following the LEAF configuration~\cite{caldas2019leaf}.

When executing both benchmarks, we observed that their execution times were smaller than the one of the TIL application. Because of that, there was no revocation with the first termination rate. Thus, we created a third revocation rate, $k_r$ being 1 hour, \textit{i.e.}, (iii) $k_r = 3600$ and $\lambda = 1/3600$.

Their execution using only on-demand VMs yields the following results. The Shakespeare application executes for almost 1 hour and 54 minutes (1:53:54) having a total cost of \$53.31, while the FEMNIST application executes for 1 hour and 56 minutes (1:56:37) with a total cost of \$35.68. The former costs more than the latter because it has 8 clients and the FEMNIST one has only 5 clients, despite using the same VMs for server and clients.

Tables \ref{tab:failure_simulation_shakespeare_bench} and \ref{tab:failure_simulation_femnist_bench}  show the results for the Shakespeare and FEMNIST applications in CloudLab with the two previous scenarios under the two termination rates: 1$/$(1 hour) and 1$/$(2 hours). Both tables consider the average values for 3 executions.

\begin{table}[h!tpp]
\caption{Failure simulation using Shakespeare changing to the same VM in CloudLab} \label{tab:failure_simulation_shakespeare_bench}
\scriptsize
\begin{center}
\begin{tabular}{|c|c|r|r|r|}
\hline
Scenario                      & \begin{tabular}[c]{@{}c@{}}Termination\\ rate\end{tabular} & \multicolumn{1}{|c|}{\begin{tabular}[c]{@{}c@{}}Avg \#\\ revoc.\end{tabular}} & \multicolumn{1}{c|}{\begin{tabular}[c]{@{}c@{}}Avg exec.\\ time\end{tabular}} & \multicolumn{1}{c|}{\begin{tabular}[c]{@{}c@{}}Avg total\\ costs\end{tabular}} \\
\hline
\multirow{2}{*}{Server and clients on spot VMs} & $k_r = 3600$ & 1.33 & 2:17:12 & \$20.02 \\
\cline{2-5}
                              & $k_r = 7200$ & 0.00 & 1:58:31 & \$17.03 \\
\hline
\multirow{2}{*}{\begin{tabular}[c]{@{}c@{}}Server on an on-demand VM\\and clients on spot VMs\end{tabular}} & $k_r = 3600$ & 2.67 & 2:32:12 & \$23.46 \\
\cline{2-5}
                              & $k_r = 7200$ & 0.00 & 1:57:56 & \$17.27 \\
\hline
% \multirow{2}{*}{\begin{tabular}[c]{@{}c@{}}Server on a spot VM\\and clients on on-demand VMs\end{tabular}}  & $k_r = 3600$ & 0.33 & 2:00:47 & \$57.06 \\
% \cline{2-5}
%                               & $k_r = 7200$ & 0.00 & 1:54:06 & \$53.29 \\
% \hline
\end{tabular}
\end{center}
\end{table}
\begin{table}[h!tpp]
\caption{Failure simulation using FEMNIST changing to the same VM in CloudLab} \label{tab:failure_simulation_femnist_bench}
\scriptsize
\begin{center}
\begin{tabular}{|c|c|r|r|r|}
\hline
Scenario                      & \begin{tabular}[c]{@{}c@{}}Termination\\ rate\end{tabular} & \multicolumn{1}{|c|}{\begin{tabular}[c]{@{}c@{}}Avg \#\\ revoc.\end{tabular}} & \multicolumn{1}{c|}{\begin{tabular}[c]{@{}c@{}}Avg exec.\\ time\end{tabular}} & \multicolumn{1}{c|}{\begin{tabular}[c]{@{}c@{}}Avg total\\ costs\end{tabular}} \\
\hline
\multirow{2}{*}{Server and clients on spot VMs} & $k_r = 3600$ & 2.00 & 2:34:33 & \$14.63 \\
\cline{2-5}
                              & $k_r = 7200$ & 0.00 & 1:52:21 & \$10.21 \\
\hline
\multirow{2}{*}{\begin{tabular}[c]{@{}c@{}}Server on an on-demand VM\\and clients on spot VMs\end{tabular}} & $k_r = 3600$ & 1.67 & 2:38:05 & \$16.10 \\
\cline{2-5}
                              & $k_r = 7200$ & 0.00 & 1:56:02 & \$11.35 \\
\hline
% \multirow{2}{*}{\begin{tabular}[c]{@{}c@{}}Server on a spot VM\\and clients on on-demand VMs\end{tabular}}  & $k_r = 3600$ & 0.67 & 2:14:55 & \$42.15 \\
% \cline{2-5}
%                               & $k_r = 7200$ & 0.00 & 1:51:07 & \$33.10 \\
% \hline
\end{tabular}
\end{center}
\end{table}

The three executions of the Shakespeare application in the first scenario with the first termination rate revoked only clients: two of them revoked only one client and the last one revoked 2 clients, resulting in an average of 1.33 revocations. For FEMNIST, two executions revoked a client and the server (2 tasks in total each) and the last one revoked 2 clients, resulting in the average of 2 revocations. 

Similarly to the TIL application, we can observe that the use of spot instances with these benchmarks is an advantage, especially to the clients tasks. 
Regarding the server execution in an on-demand VM vs. all tasks in spot VMs, costs increase on both 
%Shakespeare and FEMNIST
applications, independently of having or not revocations. In the Shakespeare application, we observe that the increase is only by 1.39\% when there is no revocation (\$17.27 vs. \$17.03) and by 17.16\% when there are revocations (\$23.46 vs. \$20.02). In the FEMNIST application, both increases are similar, 11.12\% without revocations (\$11.15 vs. \$10.21) and 10.10\% with revocations (\$16.10 and \$ 14.63).

\subsection{Execution in AWS/GCP}

In this section, we present a proof-of-concept that our framework works in a real-world multi-cloud scenario. The real-world application TIL was run on both Amazon Web Services (AWS) and Google Cloud Provider (GCP), using three regions in total: region \textit{us-east-1} (N. Virginia) in AWS and regions \textit{us-central1} (Iowa) and \textit{us-west1} (Oregon) in GCP. The instances selected per region are the same as in previous work~\cite{brum2022sbac}, as shown in Table \ref{tab:setup_poc}. 
The \textit{g4dn.2xlarge} instance has a Nvidia Tesla T4 Tensor Core GPU with 16 GB of memory while the \textit{g3.4xlarge} instance has a Nvidia Tesla M60 with 8 GB of memory. In GCP, the Turing GPU is the same as the \textit{g4dn.2xlarge} instance
% , the Pascal GPU is a Nvidia Tesla P4 witgh 8 GB of memory 
and the Volta GPU is an Nvidia V100 Tensor Core, with 16 GB of memory.
\begin{table}[h!tpp]
\caption{Instance types selected} \label{tab:setup_poc}
\tiny
\begin{center}
\begin{tabular}{|c|c|l|c|c|c|c|c|}
\hline
\multirow{2}{*}{Cloud} & \multirow{2}{*}{Region} & \multirow{2}{*}{VM} & \multirow{2}{*}{vCPUS} & \multirow{2}{*}{\begin{tabular}[c]{@{}c@{}}RAM\\(GB)\end{tabular}} & \multicolumn{2}{c|}{\begin{tabular}[c]{@{}c@{}}Costs per hour (\$)\end{tabular}} & \multirow{2}{*}{ID} \\ 
\cline{6-7}
 & & & & & On-demand & Spot & \\
\hline
\multirow{3}{*}{AWS} & \multirow{3}{*}{\begin{tabular}[c]{@{}c@{}}N. Virginia\\ (us-east-1)\end{tabular}} & \textit{g4dn.2xlarge} & 8 & 32 & 0.752 & 0.318 & $vm_{311}$ \\
\cline{3-7} 
 & & \textit{g3.4xlarge} & 16 & 122 & 1.140 & 0.638 & $vm_{312}$ \\ 
\cline{3-7}
 & & \textit{t2.xlarge} & 4 & 16 & 0.186 & 0.140 & $vm_{313}$ \\
\hline
\multirow{5}{*}{GCP} & \multirow{3}{*}{\begin{tabular}[c]{@{}c@{}}Iowa\\ (us-central1)\end{tabular}} & \textit{n1-standard-8} with Turing GPU & 8 & 30 & 0.730 & 0.196 & $vm_{411}$ \\ 
% \cline{3-7} 
%  & & \textit{n1-standard-16} with Pascal GPU & 16 & 60 & 1.360 & 0.398 & $vm_{412}$ \\
\cline{3-7} 
 & & \textit{n1-standard-8} with Volta GPU & 8 & 30 & 2.860 & 0.857 & $vm_{413}$ \\
\cline{3-7}
 & & \textit{e2-standard-4} & 4 & 16 & 0.134 & 0.040 & $vm_{414}$ \\
\cline{2-7}
 & \multirow{2}{*}{\begin{tabular}[c]{@{}c@{}}Oregon\\ (us-west1)\end{tabular}} & \textit{n1-standard-8} with Volta GPU & 8 & 30 & 2.860 & 0.857 & $vm_{422}$ \\
\cline{3-7}
 & & \textit{e2-standard-4} & 4 & 16 & 0.134 & 0.040 & $vm_{423}$ \\
\hline
\end{tabular}
\end{center}
\end{table}

In this experiment, due to GPU and vCPU restrictions, we only considered 2 clients whose dataset is stored in AWS and the GCP respectively. Our Initial Mapping module computed the optimal setup as all tasks running in AWS, with the server in VM $vm_{313}$ and the clients in VMs $vm_{311}$. The average values of three executions in only on-demand VMs yielded a runtime of 2:00:18 and a cost of \$3.28. When FL was executed with all tasks in Spot VMs and failures were simulated with a termination rate of 2 hours ($k_r = 7200$), we obtained an average number of revocations of 1.33, i.e., 1 server revocation in each run and 1 client revocation in one run. The average execution time of these runs was 2:06:51 and the costs are \$1.41.

% We execute our framework in a real cloud scenario as proof of concept using the same environment setup used in~\cite{brum2022sbac}.

% --> aqui fala dos limites

% --> Coloca as VMs da AWS/GCP aqui ou lá no environment?

% \begin{table}[h!tpp]
% \caption{Failure simulation using TIL application in AWS and GCP} \label{tab:failure_AWSGCP}
% \small
% \begin{center}
% \begin{tabular}{|c|c|r|r|r|}
% \hline
% Scenario                      & \begin{tabular}[c]{@{}c@{}}Termination\\ rate\end{tabular} & \multicolumn{1}{|c|}{\begin{tabular}[c]{@{}c@{}}Avg \#\\ revoc.\end{tabular}} & \multicolumn{1}{c|}{\begin{tabular}[c]{@{}c@{}}Avg exec.\\ time\end{tabular}} & \multicolumn{1}{c|}{\begin{tabular}[c]{@{}c@{}}Avg total\\ costs\end{tabular}} \\
% \hline
% \multirow{2}{*}{Server and clients on spot VMs} & $k_r = 7200$ & & & \\
% \cline{2-5}
%                               & $k_r = 14400$ & & & \\
% \hline
% \end{tabular}
% \end{center}
% \end{table}

\section{Conclusions and Future Work} \label{sec:conclusion}

We presented in this article the Multi-FedLS, a multi-cloud framework for Cross-Silo FL applications, which aims at reducing execution time and financial costs of such applications.
%Multi-FedLS manages multi-cloud resources of Cross-Silo FL applications. 
Multi-FedLS can be composed of both on-demand and preemptible VMs, providing a checkpoint mechanism and a dynamic scheduler that manage the replacement of preemptible VMs, in case of revocation.
%This framework implements a checkpoint mechanism and a dynamic scheduler for choosing VMs to replace the revoked one and restarting the corresponding faulty task reducing execution time and financial costs. 
We validated Multi-FedLS  in CloudLab~\cite{cloudlab_paper}, a platform that simulates cloud environments, with one real-world application and two FL benchmarks from LEAF~\cite{caldas2019leaf}.
We also presented a proof of concept by executing the real-world application on a multi-cloud platform with two real cloud providers, Amazon Web Services and Google Cloud Provider.
Our framework obtained a reduction cost of 56.92\% when compared to the execution of only on-demand VMs with an execution time increase of only 5.44\% in commercial clouds. 

% Multi-FedLS yields a cost saving of XXX\% comparing to the execution of a real-world application 
% E???? o que isso tem de important? Porq publicar esse paper? Quais foram os resultados que tornam esse paper publicavel em um journal?

Future research directions include the extension of Multi-FedLS for executing several FL applications simultaneously as well as the study of the impact of other models used in FL application, such as linear regression or Support Machine Vectors (SVMs).



\bibliographystyle{elsarticle-num} 
\bibliography{references}
\end{document}